# Wavelength-selective thermal nonreciprocity barely improves sky radiative cooling

Zihe Chen [1, #], Shilv Yu [1, #], Jinlong Ma [1], Bin Xie [2], Sun-Kyung Kim [3, *], Run Hu [1, 3, 4*]

[1] School of Energy and Power Engineering, Huazhong University of Science and Technology, Wuhan 430074, China

[2] School of Mechanical Science and Engineering, Huazhong University of Science and Technology, Wuhan 430074, China

[3] Department of Applied Physics, Kyung Hee University, Yongin-si, Gyeonggi-do 17104, Republic of Korea

[4] Wuhan National Laboratory for Optoelectronics, Huazhong University of Science and Technology, Wuhan 430074, China

*Email: hurun@hust.edu.cn; sunkim@khu.ac.kr;

[#] These authors contributed equally to this work.

**Abstract**:Radiative cooling has showcased great potential for passive refrigeration without extra energy consumption, while its cooling power and efficiency is confined by Kirchhoff's law, that is, the emissivity is equal to the absorptivity. The recent development of thermal nonreciprocity that breaks the limitations of Kirchhoff's law, especially in broadband manner, makes nonreciprocal radiative cooling (NRC) possible. Since there lacks of reports of NRC theoretically or experimentally, it is time to evaluate the feasibility and worthiness of develop NRC. Here, we discussed the effects of NRC at around room temperature (298.15 K) from three perspectives: ideal selective radiators, non-selective radiators, and colored radiators. Contrary to intuition, the introduction of thermal nonreciprocity in the atmospheric window (8-13 μm) only leads to a negative gain. Additionally, it should be noted that the radiators discussed in this work are horizontally placed without the influence of asymmetric external heat sources. The current findings shatter the inherent notion of NRC and offer some theoretical support for the practical realization and application of nonreciprocal radiative refrigeration devices.

## 1. Introduction

Sky radiative cooling (RC) has garnered increasing attentions in both academics and industry as an emerging zero-energy input cooling technology that continuously radiates thermal energy into the low-temperature outer space mainly through the atmospheric window (8-13 μm), thereby reducing the temperature of objects passively [1, 2]. RC has broad application prospects in many fields such as thermal management and building energy conservation [3, 4], just name a few. To achieve sky RC, various materials and structures have been proposed, such as photonic crystals [5], films [6], coating [7], wood [8], ceramics [9], fibers [10], metasurfaces [11], and metamaterials [12], and most of them regulates the reflectivity and emissivity spectra such as to achieve high reflectivity in the solar band and high emissivity in the atmospheric window, which is the basic rule for spectral regulation for sky RC, as shown in Fig. 1(a). At present, two types of radiative coolers have been extensively discussed: one with high emissivity across the entire mid-infrared band (Non-Selective), and the other with selective high emissivity only within the atmospheric window (Selective) [13]. With wider emission bandwidth, the former emits more energy but suffers from more atmosphere absorption than the latter due to the equal absorptivity and emissivity according to Kirchhoff's law [14]. Such equal relationship between the absorptivity and emissivity, $\epsilon(\theta,\lambda) = \alpha(\theta,\lambda)$, as described by Kirchhoff's law, is also called as the reciprocal principle of thermal radiation, which results in inevitable thermal absorption from the environment and kind of deteriorate the RC performance.

Fortunately, such severe requirement of Kirchhoff's law can be broken by experimentally applying magneto-optical (MO) material or Weyl semimetal without violating the laws of thermodynamics [15-25], i.e., $\epsilon(\theta,\lambda) \neq \alpha(\theta,\lambda)$. Such thermal nonreciprocity affirmatively provides new degree of freedom in thermal radiation control and enables more flexible regulation technologies in the field of thermal energy management. For example, the conventional reciprocal single-junction photovoltaic (PV) efficiency limit is the Shockley–Queisser limit (33% for single-junction silicon PV cell) [26, 27]. In order to break this limit, the configuration of multi-junction reciprocal PV cells has been proposed, but the maximum efficiency can only approach the multicolor limit of 86.8% due to the reciprocity. The introduction of thermal nonreciprocal can further break the efficiency limit of PV cells, reaching the Landsberg limit of 93.3% [28]. In addition, thermal nonreciprocity has also shown a positive role in the field of solar thermal photovoltaics [29] and thermal photovoltaics [30].

As both sky RC and thermal nonreciprocity are fundamentally and technologically important [31-33], there comes one open question that will thermal nonreciprocity improve RC performance just like in the field of photovoltaics? Intuitively, thermal nonreciprocity helps the improvement of sky RC due to the decrease of atmospheric absorption. However, due to the mechanism of thermal nonreciprocity and energy conservation principle [34], $\epsilon(\theta) = 1$ and $\alpha(\theta) = 0$, as shown in Fig. S1, can be achieved only within one half of the incident angle range, while $\epsilon(-\theta) = 0$ and $\alpha(-\theta) = 1$ will be caused at the other half of the incident angle range. As a result, it is difficult to give a direct conclusion of whether or not thermal nonreciprocity will benefit RC. As far as we know, there are few studies focusing on the comprehensive evaluation of benefit of combining thermal non-reciprocity and sky RC.

In this work, we introduce the concept of thermal nonreciprocity into ideal selective radiator (SR), non-selective radiator (NSR), and even colored radiator (CR) respectively, and discuss how thermal nonreciprocity affects the sky RC in different nonreciprocal bands. Detailed NRC modeling and discussion of the sky RC performance with and without thermal nonreciprocity are presented.

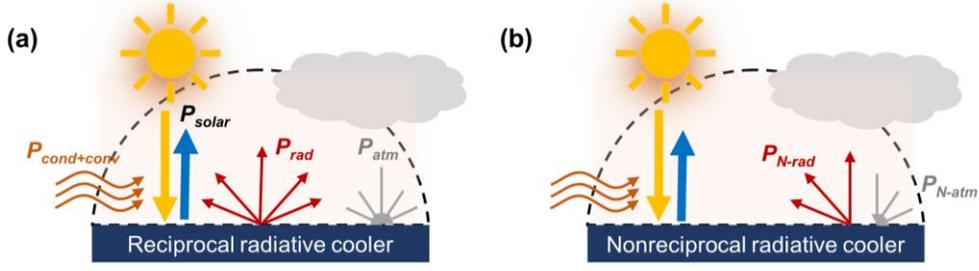

Fig. 1 Schematics of (a) reciprocal radiative cooling and (b) nonreciprocal radiative cooling. For reciprocal radiative cooler, external radiation and atmospheric environment absorption are carried out in hemispherical space. For nonreciprocal radiative cooler, in the nonreciprocal band, there is only half of the external radiation and half of the atmospheric absorption when $\epsilon(\theta)=1$ and $\alpha(\theta)=0$.

## 2. Theoretical calculation model of radiative cooling

Considering the radiator has an area $A$ and a temperature $T_e$, facing the sky with the normal direction towards the zenith. In addition, the complex environment of the cooler is simplified, only considering the standard sun and breeze environment. With such setup, its net cooling power $P_{net}$ can be described as [5, 35]:

$$P_{net}(T_e) = P_{rad}(T_e) - P_{atm}(T_{amb}) - P_{solar} - P_{cond+conv}(T_e) \tag{1}$$

where $P_{rad}$ is the power emitted from the radiator, $P_{atm}$ is the is the input power from the atmosphere absorbed by the radiator, $P_{solar}$ is the incident solar power absorbed by the radiator, and the power of the non-radiative heat transfer due to the conductive and convective is described by the $P_{cond+conv}$. $T_e$ and $T_{amb}$ are the temperature of the radiator and the ambient air, respectively. $P_{rad}$ is given by:

$$P_{rad}(T_e) = A \int d\Omega \int_0^\infty I_{BB}(T_e, \lambda) \epsilon(\lambda, \theta) \cos\theta \, d\lambda \tag{2}$$

where $\epsilon(\lambda, \theta)$ is the emissivity of the radiator at wavelength $\lambda$ and angle $\theta$. $\int d\Omega = \int_0^{\pi/2} d\theta \sin\theta \int_0^{2\pi} d\varphi$ is the solid angle integration over a hemisphere and $I_{BB}(T_e, \lambda)$ is the spectral radiance of a blackbody at temperature $T_e$ and wavelength $\lambda$, which is given by the Planck's law:

$$I_{BB}(\lambda, T_e) = \frac{2h_P c^2}{\lambda^5} \frac{1}{\exp(h_P c / \lambda k_B T_e) - 1} \tag{3}$$

where $h_P$ is the Planck's constant, $k_B$ is the Boltzmann constant and $c$ is the speed of the light. The input power from the atmosphere radiation in Eq. (1) is given by:

$$P_{atm}(T_{amb}) = A \int d\Omega \int_0^\infty I_{BB}(T_{amb}, \lambda) \alpha(\lambda, \theta) \epsilon_{atm}(\lambda, \theta) \cos\theta \, d\lambda \tag{4}$$

where $\alpha(\lambda, \theta)$ is the absorption of the radiator at wavelength $\lambda$ and angle $\theta$. $\epsilon_{atm}(\lambda, \theta)$ is the emissivity of the atmosphere, which can be calculated as: $\epsilon_{atm}(\lambda, \theta) = 1 - \tau(\lambda)^{1/\cos\theta}$, here $\tau(\lambda)$ is the transmittance of the atmosphere in the zenith direction. The input solar power is given by:

$$P_{solar} = A \cdot G \int_0^\infty \epsilon(\lambda, 0) I_{AM1.5}(\lambda) \Big/ \int_0^\infty I_{AM1.5}(\lambda) d\lambda \tag{5}$$

where $I_{AM.15}(\lambda)$ is the standard AM 1.5 spectrum of solar radiation, and $G$ is the total solar irradiance at 1 KW/m². The $P_{cond+conv}$ is given by:

$$P_{cond+conv} = A \cdot h(T_{amb} - T_e) \tag{6}$$

where $h$ is a non-radiative heat transfer coefficient that combines the effective conductive and convective heat exchange.

For the nonreciprocal radiator in Fig. 1(b), when the incidence angle is 0°, there is no nonreciprocity phenomenon, so the introduction of nonreciprocity does not affect the input solar power and the power of the non-radiative heat transfer, but only affects the power emitted from the radiator and the absorbed power from the atmosphere. Here, considering the nonreciprocal band from $\lambda_1$ to $\lambda_2$, the power emitted from the radiator under the thermal nonreciprocity $P_{N\text{-}rad}$ is given by:

$$P_{N\text{-}rad}(T_e) = \frac{1}{2} P_{rad}(T_e) = \frac{A}{2} \int d\Omega \int_{\lambda_1}^{\lambda_2} I_{BB}(T_e, \lambda) \epsilon(\lambda, \theta) \cos\theta d\lambda . \tag{7}$$

The input power from the atmosphere radiation under the thermal nonreciprocity $P_{N\text{-}atm}$ is given by:

$$P_{N\text{-}atm}(T_{amb}) = \frac{1}{2} P_{atm}(T_{amb}) = \frac{A}{2} \int d\Omega \int_{\lambda_1}^{\lambda_2} I_{BB}(T_{amb}, \lambda) \alpha(\lambda, \theta) \epsilon_{atm}(\lambda, \theta) \cos\theta d\lambda . \tag{8}$$

By integrating the above equations separately, we can obtain the net cooling power $P_{net}$ of the radiator at different temperature $T_e$. When the radiator reaches a thermal equilibrium state, the $P_{net}$ is zero, and the corresponding steady-state temperature $T_s$ can be obtained. A lower $T_s$ indicates a better cooling performance. In the following calculations, without additional explanation, the ambient temperature $T_{amb}$ is set as 298.15 K to simulate a sunny and breezy situation.

## 3. Theoretical calculation of nonreciprocal radiative cooling

In this section, we will discuss the influence of thermal nonreciprocity on RC from three aspects: 1) The effect of thermal nonreciprocity on the ideal selective radiator (R-8-13); 2) The effect of thermal nonreciprocity on the ideal non-selective radiator (R-2.5-25); 3) Effect of thermal nonreciprocity on color selective/non-selective radiator. Each section considers the influence of different nonreciprocal bands on RC and names the different bands N-4-8, N-8-13, and N-13-25, where the letters N and R represent nonreciprocal and reciprocal, respectively, and the numbers represent the band range.

### 3.1 Effect of thermal nonreciprocity on the ideal selective radiator

The theoretical emission and absorption spectra of the ideal selective radiator and their corresponding nonreciprocal radiators with different nonreciprocal bands are shown in the Fig. 2(a-d). Fig. 2(a) shows the absorptivity and emissivity spectra of the ideal selective radiator, which has a unit absorptivity/emissivity in the atmospheric window. Fig. 2(b-d) show the absorptivity and emissivity spectra of the nonreciprocal selective radiators with different nonreciprocal bands. For example, when the nonreciprocal band is 4 to 8 μm, the emissivity is 1 and the absorptivity is 0 in a half of the hemisphere, and the absorptivity is 1 and the emissivity is 0 in the other half of the hemisphere space, refer to Fig. 1 (b) and Fig. S1. Fig. 2 (e-g) discuss the effect of different nonreciprocal bands on RC with different $h$. When $h$=0 W/m$^2$/K, R-8-13 has the lowest $T_s$ of about 236.15 K, and the introduction of thermal nonreciprocity in different bands cannot reduce $T_s$, but increase $T_s$. When $h$=3 W/m$^2$/K, $T_s$ of R-8-13 is 278.12 K and that of N-4-8 is 277.88 K. The latter has a 0.24 K reduction and a higher $P_{net}$ than R-8-13, indicating that the introduction of thermal nonreciprocity in the band of 4-8 μm can slightly help RC. However, $T_s$ of N-8-13 is equal to 285.7 K, which is higher than that of R-8-13 and shows the weakest cooling performance. As $h$ continues to increase to 12 W/m$^2$/K, N-4-8 also has the best cooling performance, but even if $h$ increases to 12 W/m$^2$/K, $T_s$ can only be reduced by about 0.6 K compared with R-8-13, as shown in Fig. 2(g). In addition, $T_s$ of N-13-25 is also slightly lower than that of R-8-13 (about 0.1 K), showing a very limited gain effect. Therefore, for ideal selective radiators, the introduction of thermal nonreciprocity in different bands has limited and even harmful effects on radiative cooling.

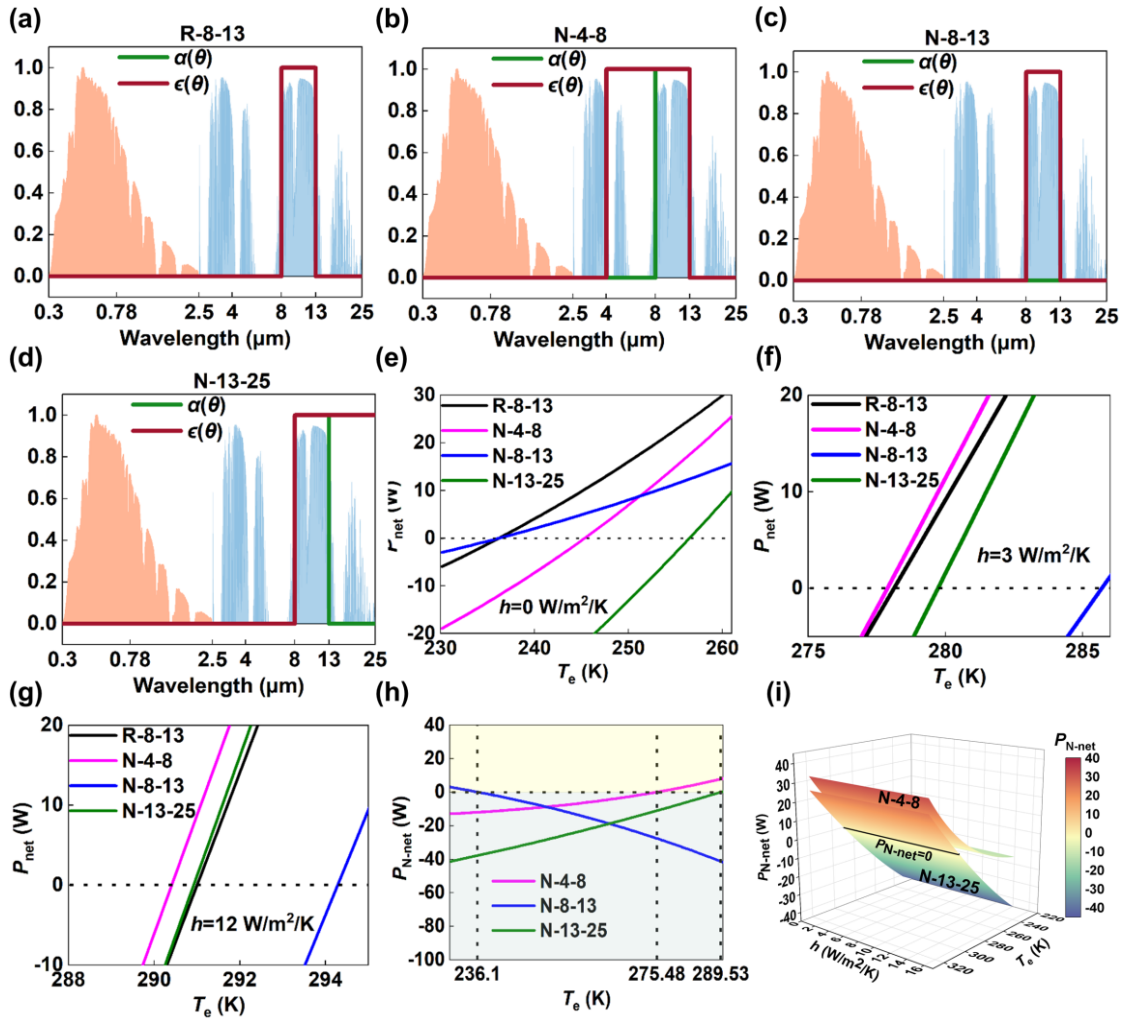

Fig.2 Ideal spectral absorptivity ($α(θ)$) and emissivity ($ε(θ)$) of the selective radiator with $α(θ)=ε(θ)=1$ and corresponding nonreciprocal radiators with $α(θ)=0$ and $ε(θ)=1$. (a) R-8-13 with unit emissivity and absorptivity in the band (8-13 μm). (b) N-4-8 with unit emissivity and zero absorptivity of in the band (4-8 μm). (c) N-8-13 with unit emissivity and zero absorptivity of in the band (8-13 μm). (d) N-13-25 with unit emissivity and zero absorptivity of in the band (13-25 μm). (e) $P_{net}$ of the selective radiator and corresponding nonreciprocal radiators with $h$=0

W/m²/K. (f) $P_{net}$ of the selective radiator and corresponding nonreciprocal radiators with $h$=3 W/m²/K. (g) $P_{net}$ of the selective radiator and corresponding nonreciprocal radiators with $h$=12 W/m²/K. (h) The net power resulting from nonreciprocity with different nonreciprocal bands. (i) Relationship between $P_{N-net}$ and $h$ of the nonreciprocal radiators (N-4-8 and N-13-25).

Here, the influence mechanism of thermal nonreciprocity on RC is analyzed by calculating the net power brought by nonreciprocity. Compared to R-8-13, when the nonreciprocal band is not 8-13 μm, the net power resulting from nonreciprocity $P_{N-net}$ is:

$$P_{N-net}(T_e) = P_{N-rad}(T_e) - P_{N-atm}(T_{amb}). \tag{9}$$

When the nonreciprocal band is 8-13 μm, the net power resulting from nonreciprocity $P_{N-net}$ is:

$$P_{N-net}(T_e) = P_{N-atm}(T_{amb}) - P_{N-rad}(T_e). \tag{10}$$

According to Eq. (9-10), the relationship between $P_{N-net}$ and $T_e$ for different nonreciprocal bands are shown in Fig. 2 (h). For the case of N-4-8, $P_{N-net}$ is positive when $T_e$ > 275.48 K, representing that thermal nonreciprocity can help RC, which explains the higher $P_{net}$ and lower $T_s$ of N-4-8 in Fig. 2 (f-g). Similarly, when $T_e$ > 289.53 K, N-13-25 can also improve the cooling performance. However, for the case of N-8-13, $P_{N-net}$ is negative when $T_e$ > 236.1 K, which shows that the introduction of thermal nonreciprocity in 8-13 μm cannot help RC. The relationship between $P_{N-net}$ and $h$ is further discussed, as shown in Fig. 2(i). For example, when $P_{N-net}$ =0, it is a horizontal line, meaning that $P_{N-net}$ does not change with $h$.

To sum up, in order to more clearly show the effect of thermal nonreciprocity on the selective radiator, it is summarized in Table 1. From the perspective of power gain, when $T_e$ is higher than 275.48 K/289.53 K, the introduction of thermal nonreciprocity in the band of 4-8 μm /13-25 μm can help RC, corresponding to Fig. 2(h). From the perspective of equilibrium temperature, for example, when $h$=12 W/m²/K, N-4-8/N-13-25 can achieve the reduction of $T_s$, but the reduction degree is only 0.6K/0.1K, corresponding to Fig. 2(g), showing a very limited gain effect. In addition, we note that when $T_e$ > 236.1 K, the introduction of thermal nonreciprocity in atmospheric window only lead to a reduced radiative cooling effect.

**Table 1** Effect of different nonreciprocal bands on SR ($T_{amb}$ = 298.15 K)

| Cases | Cooling power gain | $\Delta T_s$ ($h$=12 W/m²/K) |
|---|---|---|
| N-4-8 | Positive gain, $T_e$ > 275.48 K | $\Delta T_s$ = 0.6 K |
| N-8-13 | Negative gain, $T_e$ > 236.1 K | $\Delta T_s$ = -3.3 K |
| N-13-25 | Positive gain, $T_e$ > 289.53 K | $\Delta T_s$ = 0.1 K |

## 3.2 Effect of thermal nonreciprocity on the ideal non-selective radiator

Next, we investigate the effect of thermal nonreciprocity on non-selective radiators. The theoretical emission and absorption spectra of the ideal non-selective radiator and the corresponding nonreciprocal radiators with different nonreciprocal bands are shown in the Fig. 3(a-d). Fig. 3(a) shows the absorptivity and emissivity spectra of the ideal non-selective radiator, which has a unit absorptivity/emissivity in the band of 2.5-25 μm. Fig. 3(b-d) show the absorptivity and emissivity spectra of the nonreciprocal non-selective radiators with different nonreciprocal bands in half of hemispherical space. Fig. 3 (e-g) discuss the effect of different nonreciprocal bands on RC with different $h$. When $h = 0$ W/m²/K, N-13-25 has the lowest $T_s$ of about 264.4 K, which is about 5 K lower than that of R-2.5-25 (269.4 K), indicating that the introduction of thermal nonreciprocity in 13-25μm can help RC for non-selective radiators. In addition, $T_s$ of N-4-8 is about 268.65 K, which is about 0.75 K lower than that of R-2.5-25, which also shows a certain gain effect. However, for the case of N-8-13, $T_s$ is 276.2 K, showing the weakest cooling performance compared with other radiators. When $h = 3$ W/m²/K, as shown in Fig.3(f), only N-13-25 still has a better cooling performance than R-2.5-25, which is about 1 K lower than that of R-2.5-25. As $h$ increases to 12 W/m²/K, the thermal nonreciprocity hardly helps RC for the non-selective radiator, as shown in Fig. 3(g).

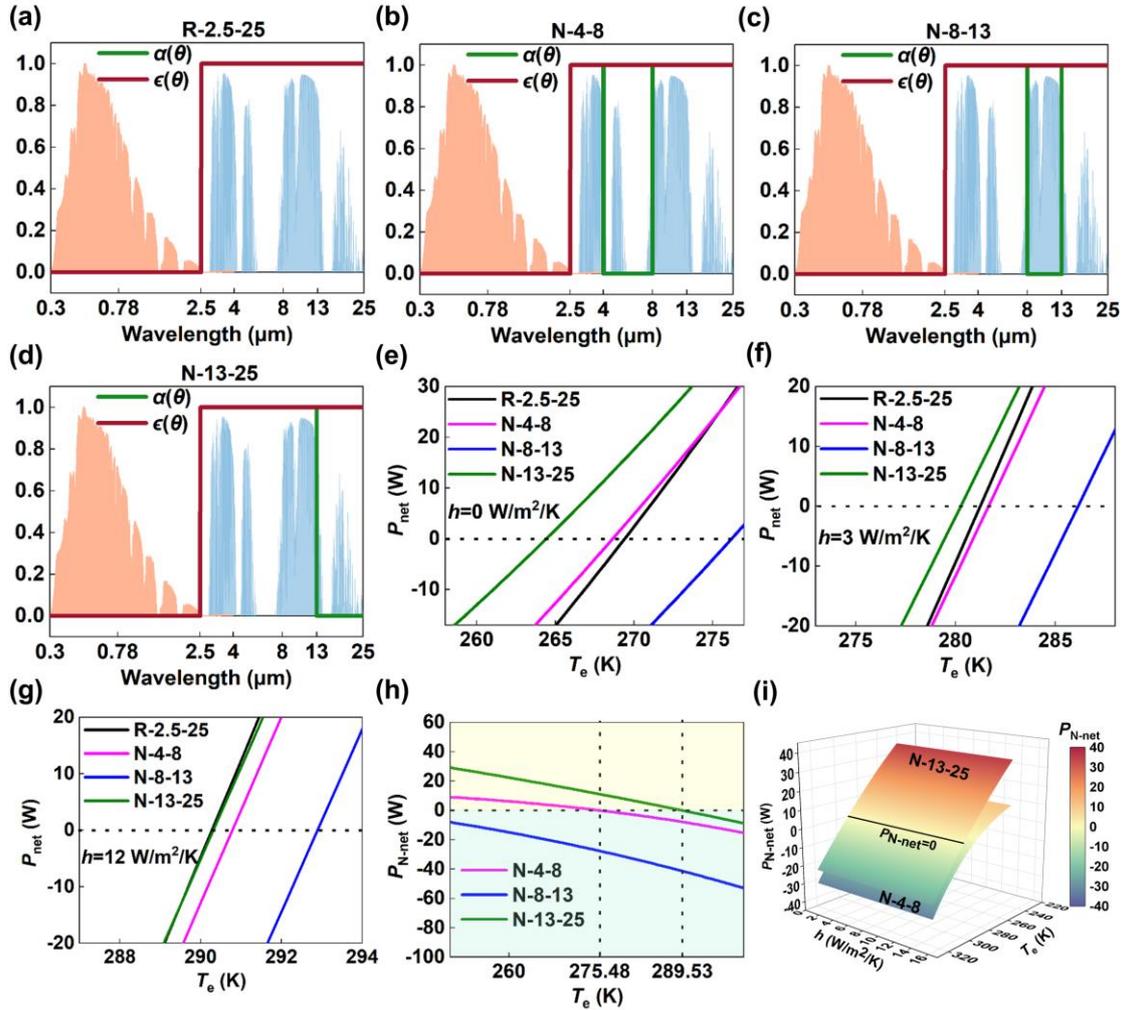

Fig.3 Ideal spectral absorptivity ($\alpha(\theta)$) and emissivity ($\epsilon(\theta)$) of the non-selective radiator with $\alpha(\theta)=\epsilon(\theta)=1$ and corresponding nonreciprocal radiators with $\alpha(\theta)=0$ and $\epsilon(\theta)=1$. (a) R-2.5-25 with unit emissivity and absorptivity in the band (8-13 μm). (b) N-4-8 with unit emissivity and zero absorptivity of in the band (4-8 μm). (c) N-8-13 with unit emissivity and zero absorptivity of in the band (8-13 μm). (d) N-13-25 with unit emissivity and zero absorptivity of in the band (13-25 μm). (e) $P_{net}$ of the non-selective radiator and corresponding nonreciprocal radiators with $h$=0 W/m²/K. (f) $P_{net}$ of the non-selective radiator and corresponding nonreciprocal radiators with $h$=3 W/m²/K. (g) $P_{net}$ of the non-selective radiator and corresponding nonreciprocal radiators with $h$=12 W/m²/K. (h) The net power

resulting from nonreciprocity with different nonreciprocal bands. (i) Relationship between $P_{\text{N-net}}$ and $h$ of the nonreciprocal radiators (N-4-8 and N-13-25).

Here, the influence mechanism of thermal nonreciprocity on R-2.5-25 is analyzed by calculating the net power brought by thermal nonreciprocity. Compared to R-2.5-25, the net power resulting from nonreciprocity $P_{\text{N-net}}$ is:

$$P_{\text{N-net}}(T_e) = P_{\text{N-atm}}(T_{\text{amb}}) - P_{\text{N-rad}}(T_e). \tag{11}$$

According to Eq. (11), the changes of net power with $T_e$ in different nonreciprocal bands are shown in Fig. 3 (h). For the case of N-4-8, $P_{\text{N-net}}$ is positive when $T_e < 275.48$ K, representing that thermal nonreciprocity can help RC, which also explains the lower $T_s$ of N-4-8 than that of R-2.5-25 in Fig. 3 (e). Similarly, for the case of N-13-25, when $T_e < 289.53$ K, $P_{\text{N-net}}$ is positive, which explains the lower $T_s$ of N-13-25 than that of R-2.5-25 in Fig. 3 (e-f). However, for the case of N-8-13, $P_{\text{N-net}}$ is negative, which explains why atmospheric window cannot be selected as the nonreciprocal band. The relationship between $P_{\text{N-net}}$ and $h$ is also further discussed for nonreciprocal radiators, as shown in Fig. 3(i). For example, when $P_{\text{N-net}} = 0$, it is a horizontal line, meaning that $P_{\text{N-net}}$ does not change with $h$.

To sum up, in order to more clearly show the effect of thermal nonreciprocity on the non-selective radiator, it is summarized in Table 2. From the point of view of power gain, when $T_e$ is lower than 275.48 K/289.53 K, the introduction of thermal nonreciprocity in the band 4-8 μm /13-25 μm can realize a positive gain, corresponding to Fig. 3(h). From the perspective of $T_s$, N-4-8/N-13-25 can achieve the reduction of $T_s$ and the reduction degree is 0.75 K/5 K when $h=0$ W/m²/K, corresponding to Fig. 3(e). However, as $h$ gradually increases, the gain effect is gradually weakened and even negative gain. In addition, similar to section 3.1, the introduction of thermal nonreciprocity in the atmospheric window only compromises the radiative cooling effect.

**Table 2** Effect of nonreciprocal band on ideal non-selective radiator ($T_{\text{amb}} = 298.15$ K)

| Cases | Power gain | $\Delta T_s$ ($h=0$ W/m²/K) | $\Delta T_s$ ($h=3$ W/m²/K) |
|---|---|---|---|
| N-4-8 | Positive gain, $T_e < 275.48$ K | $\Delta T_s = 0.75$ K | $\Delta T_s = -0.41$ K |
| N-8-13 | Negative gain | $\Delta T_s = -6.75$ K | $\Delta T_s = -4.91$ K |
| N-13-25 | Positive gain, $T_e < 289.53$ K | $\Delta T_s = 5$ K | $\Delta T_s = 1$ K |

### 3.3 Effect of thermal nonreciprocity on colored radiators

Both reciprocal and nonreciprocal radiators discussed above face a new problem, that is, for the purpose of maximizing the RC effect, they present total reflection in the solar band, which makes the radiators in white appearance. The large area of white appearance is terrible for aesthetic requirements and resulting in potential of light pollution. Consequently, in pursuit of practicality, colored radiators (CRs) have been developed, which demonstrate rich color but discounted cooling performance [36]. From the perspective of spectrum, CR shows color because it exhibits partial reflection rather than total reflection in visible band, which results in absorption of solar radiation and thus weakens the cooling effect. Since the color and cooling performance are a kind of competing, existing reciprocal based CRs are limited in color richness, and most of them display light colors [37, 38].

So, in this section, we discuss whether the thermal nonreciprocity will bring benefit to CRs and have a profound impact. Since darker colors tend to have larger absorption of solar energy than light colors, they are difficult to achieve sub-ambient cooling in conventional reciprocal RC [39]. For this purpose, we randomly chose a deep color, namely deep magenta, to explore the effect of thermal nonreciprocity on the CRs and determine whether it can achieve better RC effect. Of course, other dark colors can also be chosen, such as gray, as shown in Fig. S3, whose conclusions are consistent with those of the dark magenta case. Fig. 4(a) displays the reflectance spectrum of magenta, which has low reflectance especially in range of 0.48~0.65 μm and results in high absorptivity (emissivity). Here, we consider the effect of nonreciprocity on color non-selective radiator (C-R-2.5-25) and color selective radiator (C-R-8-13). Firstly, for C-R-2.5-25, its absorption and emission spectra are shown in Fig. 4 (b). Both the emissivity and absorption are 1 in the band range of 2.5-25 μm, and the visible light band corresponds to the deep magenta spectrum. Fig. S2 shows the relationship of $P_{net}$ of C-R-2.5-25 (deep magenta) with $T_e$ and $h$. When $P_{net}$ >0, $T_e$ is higher than 300 K, so both N-4-8 and N-13-25 cannot improve the RC effect for C-R-2.5-25, according to Table 2. Secondly, for the case of C-R-8-13, its absorption and emission spectra are shown in Fig. 4 (c). The emissivity and absorption are both 1 in the atmospheric window, and the visible light band corresponds to the deep magenta spectrum. Fig. 4(d) shows the relationship between $P_{net}$ and $h$ of C-R-8-13 with a deep magenta color. It can be seen that when $P_{net}$ >0, $T_e$ is higher than 300 K. According to Fig. 2(h) and Table 1, since $T_e$ of C-R-8-13 is higher than 289.53 K, both N-4-8 and N-13-25 can improve the RC effect and the corresponding nonreciprocal spectra are shown in Fig. 4(e). When $h$=0 W/m²/K, the net power of C-R-8-13 and the corresponding nonreciprocal radiator (C-N-4-8&13-25) changes with $T_e$, as shown in Fig. 4(f). The introduction of thermal nonreciprocity can reduce $T_s$ by 21.26 K compared with C-R-8-13, showing a better cooling performance. However, the cooling performance of C-N-4-8&13-25 is weaker than that of C-R-2.5-25. Therefore, the design of reciprocal non-selective radiators is better for CRs with a dark color. In addition, it should be noted that as the color of the radiator gradually becomes lighter or even white, the advantage of the non-selective radiator will diminish, as shown in Fig. 4(g-i) and Fig. S4. This is mainly because when the color is lighter, less solar radiation is absorbed and $T_s$ ($P_{net}$=0) is lower. As shown in Table 2, when $T_s$ gradually decreases to 289.53 K, the introduction of thermal nonreciprocity into the non-atmospheric window will gradually play a positive gain effect. In addition, with the color radiator light to white, as shown in Fig. S4, the performance of nonreciprocal radiator (N-4-8&13-25) is better than that of non-selective thermal radiator (R-2.5-25) but weaker than that of selective radiator (R-8-13), which is consistent with sections 3.1 and 3.2.

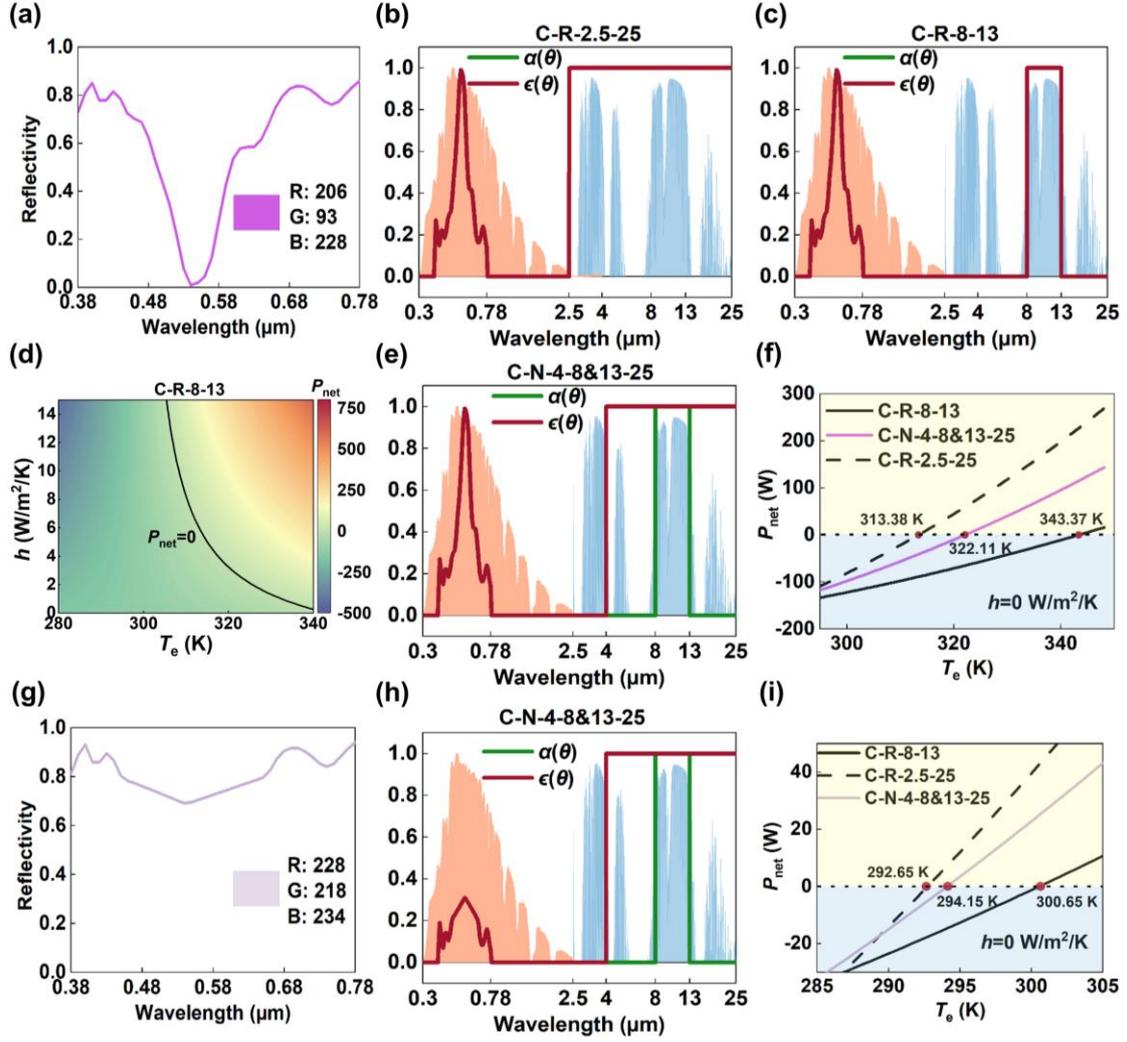

Fig.4 (a) The reflectance spectral of deep magenta. (b) Reciprocal non-selective radiator with magenta (C-R-2.5-25). (c) Reciprocal selective radiator with magenta (C-R-8-13). (d) Relationship of $P_{net}$ of C-R-8-13 with $T_e$ and $h$. (e) Nonreciprocal radiator with magenta (C-N-4-8&13-25). (f) Net cooling power of C-R-2.5-25, C-R-8-13 and C-N-4-8&13-25 with $h$=0 W/m²/K. (g) Reflectance spectrum for a light red case. (h) The spectrum of the nonreciprocal color radiator with the light red color (C-N-4-8&13-25). (i) Net cooling power of C-R-8-13, C-R-2.5-25 and C-N-4-8&13-25 with $h$=0 W/m²/K for the light red case.

## 4. Conclusion

In this work, we consider the effects of thermal nonreciprocity on ideal selective radiators, non-selective radiators, and colored radiators. For the three radiators, the introduction of thermal nonreciprocity in the atmospheric window (8-13μm) will be harmful to the cooling effect. In addition, for the selective radiator, when $T_e$ is higher than 275.48 K/289.53 K, the introduction of thermal nonreciprocity in the band of 4-8 μm/13-25 μm can help RC effect. However, even if with $h$=12 W/m²/K, the $T_s$ of N-4-8/ N-13-25 is only 0.6 K/0.1K lower than that of R-8-13, which shows a slight enhancement effect. For non-selective radiators, when $h$ = 0W/m²/K, the introduction of thermal nonreciprocity in the band of 13-25 μm can realize an obvious gain effect. However, with the increase of $h$, the gain effect due to thermal nonreciprocity becomes weaker and even negative gain. For example, when $h$=3 W/m$^2$/K, the introduction of nonreciprocity only reduces $T_s$ by about 1 K. For the color radiators with dark colors, C-R-2.5-25 can realize a better RC effect compared with nonreciprocal radiators. In addition, it should be noted that the above theoretical work is calculated under extremely ideal conditions: 1) Consider that both TM and TE waves can achieve thermal nonreciprocity (currently mainly implemented under TM waves); 2) Consider that the thermal nonreciprocity does not vary with the azimuth angle. Therefore, currently, considering the weak gain effect and high application requirements, achieving efficient nonreciprocal radiative cooling solely through the introduction of thermal nonreciprocity in different bands remains challenging.

**Author contributions**

**Zihe Chen:** Conceptualization, Methodology, Validation, Formal analysis, Data curation, Writing – original draft, Writing – review & editing. **Shilv Yu:** Conceptualization, Methodology, Validation, Formal analysis, Data curation, Writing – original draft, Writing – review & editing. **Jinlong Ma:** Supervision, Writing - review & editing. **Bin Xie:** Supervision, Writing - review & editing. **Sun-Kyung Kim:** Formal analysis, Supervision, Writing - review & editing. **Run Hu:** Conceptualization, Formal analysis, Writing – original draft, Writing – review & editing.

**Declaration of Competing Interest**

The authors declare that they have no known competing financial interests or personal relationships that could have appeared to influence the work reported in this paper.

**Data availability**

Data will be made available on request.

**Acknowledgement**

The authors would like to acknowledge the financial support by National Natural Science Foundation of China (52422603, 52076087, 52211540005), the Open Project Program of Wuhan National Laboratory for Optoelectronics (2021WNLOKF004), Interdisciplinary Research Program of HUST (5003120094), and Natural Science Foundation of Hubei Province (2023AFA072), and the Fundamental Research Funds for the Central Universities (YCJJ20242102).

**References**
[1] S. Fan, W. Li, Photonics and thermodynamics concepts in radiative cooling, Nat. Photonics, 16(3) (2022) 182-190.
[2] A.P. Raman, M.A. Anoma, L. Zhu, E. Rephaeli, S. Fan, Passive radiative cooling below ambient air temperature under direct sunlight, Nature, 515(7528) (2014) 540-544.
[3] L. Zhou, H. Song, J. Liang, M. Singer, M. Zhou, E. Stegenburgs, N. Zhang, C. Xu, T. Ng, Z. Yu, B. Ooi, Q. Gan, A polydimethylsiloxane-coated metal structure for all-day radiative cooling, Nat.

Sustainability, 2(8) (2019) 718-724.

[4] Y. Peng, L. Fan, W. Jin, Y. Ye, Z. Huang, S. Zhai, X. Luo, Y. Ma, J. Tang, J. Zhou, L.C. Greenburg, A. Majumdar, S. Fan, Y. Cui, Coloured low-emissivity films for building envelopes for year-round energy savings, Nat. Sustainability, 5(4) (2021) 339-347.

[5] M. Lee, G. Kim, Y. Jung, K.R. Pyun, J. Lee, B.W. Kim, S.H. Ko, Photonic structures in radiative cooling, Light Sci Appl, 12(1) (2023) 134.

[6] H. Yin, X. Zhou, Z. Zhou, R. Liu, X. Mo, Z. Chen, E. Yang, Z. Huang, H. Li, H. Wu, J. Zhou, Y. Long, B. Hu, Switchable Kirigami Structures as Window Envelopes for Energy-Efficient Buildings, Research, 2023(4) (2023) 0103.

[7] C. Feng, Y. Lei, X.Q. Huang, W.D. Zhang, Y. Feng, X. Zheng, Experimental and theoretical analysis of sub-ambient cooling with longwave radiative coating, Renewable Energy, 193 (2022) 634-644.

[8] T. Li, Y. Zhai, S.M. He, W.T. Gan, Z.Y. Wei, M. Heidarinejad, D. Dalgo, R.Y. Mi, X.P. Zhao, J.W. Song, J.Q. Dai, C.J. Chen, A. Aili, A. Vellore, A. Martini, R.G. Yang, J. Srebric, X.B. Yin, L.B. Hu, A radiative cooling structural material, Science, 364(6442) (2019) 760-+.

[9] K.X. Lin, S.R. Chen, Y.J. Zeng, T.C. Ho, Y.H. Zhu, X. Wang, F.Y. Liu, B.L. Huang, C.Y.H. Chao, Z.K. Wang, C.Y. Tso, Hierarchically structured passive radiative cooling ceramic with high solar reflectivity, Science, 382(6671) (2023) 691-697.

[10] X. Wang, X. Liu, Z. Li, H. Zhang, Z. Yang, H. Zhou, T. Fan, Scalable Flexible Hybrid Membranes with Photonic Structures for Daytime Radiative Cooling, Adv. Funct. Mater., 30(5) (2019) 1907562.

[11] M.M. Hossain, B.H. Jia, M. Gu, A Metamaterial Emitter for Highly Efficient Radiative Cooling, Adv. Opt. Mater., 3(8) (2015) 1047-1051.

[12] M.M. Hossain, B. Jia, M. Gu, A Metamaterial Emitter for Highly Efficient Radiative Cooling, Adv. Opt. Mater., 3(8) (2015) 1047-1051.

[13] Z.F. Huang, X.L. Ruan, Nanoparticle embedded double-layer coating for daytime radiative cooling, Int. J. Heat Mass Transfer, 104 (2017) 890-896.

[14] M.M. Hossain, M. Gu, Radiative Cooling: Principles, Progress, and Potentials, Adv. Sci., 3(7) (2016) 1500360.

[15] C. Guo, B. Zhao, S. Fan, Adjoint Kirchhoff's Law and General Symmetry Implications for All Thermal Emitters, Phys. Rev. X, 12(2) (2022) 021023.

[16] A. Ghanekar, J.H. Wang, S.H. Fan, M.L. Povinelli, Violation of Kirchhoff's Law of Thermal Radiation with Space-Time Modulated Grating, ACS Photonics, 9(4) (2022) 1157-1164.

[17] Z. Zhang, X. Wu, C. Fu, Validity of Kirchhoff's law for semitransparent films made of anisotropic materials, Journal of Quantitative Spectroscopy & Radiative Transfer, 245 (2020) 106904.

[18] Z. Chen, S. Yu, B. Hu, R. Hu, Multi-band and wide-angle nonreciprocal thermal radiation, Int. J. Heat Mass Transfer, 209 (2023) 124149.

[19] Z. Chen, S. Yu, C. Yuan, K. Hu, R. Hu, Ultra-efficient machine learning design of nonreciprocal thermal absorber for arbitrary directional and spectral radiation, J. Appl. Phys., 134(20) (2023) 203101.

[20] Z. Chen, S. Yu, C. Yuan, X. Luo, R. Hu, Near-normal nonreciprocal thermal radiation with a 0.3T magnetic field based on double-layer grating structure, Int. J. Heat Mass Transfer, 222 (2024) 125202.

[21] K. Shi, Y. Xing, Y. Sun, N. He, T. Guo, S. He, Thermal Vertical Emitter of Ultra‐High Directionality Achieved Through Nonreciprocal Magneto‐Optical Lattice Resonances, Adv. Opt. Mater., 10(24) (2022) 2201732.

[22] K.J. Shayegan, S. Biswas, B. Zhao, S. Fan, H.A. Atwater, Direct observation of the violation of Kirchhoff's law of thermal radiation, Nat. Photonics, 17 (2023) 891–896.

[23] M. Liu, S. Xia, W. Wan, J. Qin, H. Li, C. Zhao, L. Bi, C.-W. Qiu, Broadband mid-infrared non-reciprocal absorption using magnetized gradient epsilon-near-zero thin films, Nat. Mater., 22 (2023) 1196–1202.

[24] K.J. Shayegan, B. Zhao, Y. Kim, S. Fan, H.A. Atwater, Nonreciprocal infrared absorption via resonant magneto-optical coupling to InAs, Sci. Adv., 8 (2022) eabm4308

[25] K. Shi, Y. Sun, R. Hu, S. He, Ultra-broadband and wide-angle nonreciprocal thermal emitter based on Weyl semimetal metamaterials, Nanophotonics, 13(5) (2024) 737-747.

[26] M.A. Green, Time-asymmetric photovoltaics, Nano Lett., 12(11) (2012) 5985-5988.

[27] T. Lu, M.J. Li, W.C. Lu, T.Y. Zhang, Recent progress in the data-driven discovery of novel photovoltaic materials, Journal of Materials Informatics, 2(2) (2022) 1-44.

[28] Y. Park, B. Zhao, S. Fan, Reaching the Ultimate Efficiency of Solar Energy Harvesting with a Nonreciprocal Multijunction Solar Cell, Nano Lett., 22(1) (2022) 448-452.

[29] S. Jafari Ghalekohneh, B. Zhao, Nonreciprocal Solar Thermophotovoltaics, Phys. Rev. Appl., 18(3) (2022) 034083.

[30] Y. Park, Z. Omair, S. Fan, Nonreciprocal Thermophotovoltaic Systems, ACS Photonics, 9(12) (2022) 3943-3949.

[31] M.Q. Liu, C.Y. Zhao, Near-infrared nonreciprocal thermal emitters induced by asymmetric embedded eigenstates, Int. J. Heat Mass Tran., 186 (2022) 122435.

[32] J. Wu, Y.M. Qing, Tunable near-perfect nonreciprocal radiation with a Weyl semimetal and graphene, PCCP, 25(13) (2023) 9586-9591.

[33] J. Wu, B. Wu, Z. Wang, X. Wu, Strong nonreciprocal thermal radiation in Weyl semimetal-dielectric multilayer structure, Int. J. Therm. Sci., 181 (2022) 107788.

[34] L. Zhu, S. Fan, Near-complete violation of detailed balance in thermal radiation, Physical Review B, 90(22) (2014) 220301.

[35] W. Xi, Y. Liu, W. Zhao, R. Hu, X. Luo, Colored radiative cooling: How to balance color display and radiative cooling performance, Int. J. Therm. Sci., 170 (2021) 107172.

[36] B. Xie, Y.D. Liu, W. Xi, R. Hu, Colored radiative cooling: progress and prospects, Mater. Today Energy, 34 (2023) 101302.

[37] J.W. Cho, E.J. Lee, S.K. Kim, Full-Color Solar-Heat-Resistant Films Based on Nanometer Optical Coatings, Nano Lett., 22(1) (2022) 380-388.

[38] W. Xi, Y.D. Liu, W.X. Zhao, R. Hu, X.B. Luo, Colored radiative cooling: How to balance color display and radiative cooling performance, Int. J. Therm. Sci., 170 (2021) 107172.

[39] W. Li, Y. Shi, Z. Chen, S.H. Fan, Photonic thermal management of coloured objects, Nat. Commun., 9 (2018) 4240.

# Supplementary Information for

## Wavelength-selective thermal nonreciprocity barely improves sky radiative cooling

Zihe Chen [1, #], Shilv Yu [1, #], Jinlong Ma [1], Bin Xie [2], Sun-Kyung Kim [3, *], Run Hu [1, 3, 4*]

[1]School of Energy and Power Engineering, Huazhong University of Science and Technology, Wuhan 430074, China

[2]School of Mechanical Science and Engineering, Huazhong University of Science and Technology, Wuhan 430074, China

[3]Department of Applied Physics, Kyung Hee University, Yongin-si, Gyeonggi-do 17104, Republic of Korea

[4]Wuhan National Laboratory for Optoelectronics, Huazhong University of Science and Technology, Wuhan 430074, China

*Email: hurun@hust.edu.cn; sunkim@khu.ac.kr;

[#] These authors contributed equally to this work.

**Content**

**Fig. S1** (a) Energy flow diagrams in the case of a nonreciprocal radiator; (b) Schematic diagram when $\epsilon_\theta = 1\ and\ \alpha_\theta = 0$.

**Fig. S2** Relationship of $P_{net}$ with $T_e$ and $h$ for C-R-2.5-25 with a deep magenta color.

**Fig. S3** (a) Reflectance spectrum for another dark case (gray); (b) The spectrum of the nonreciprocal color radiator (C-N-4-8&13-25); (c) Net cooling power of C-R-8-13, C-R-2.5-25 and C-N-4-8&13-25 with $h$=0 W/m²/K for the gray case.

**Fig. S4** Net cooling power of R-2.5-25, R-8-13 and N-4-8&13-25 with $h$=0 W/m²/K for the white case.

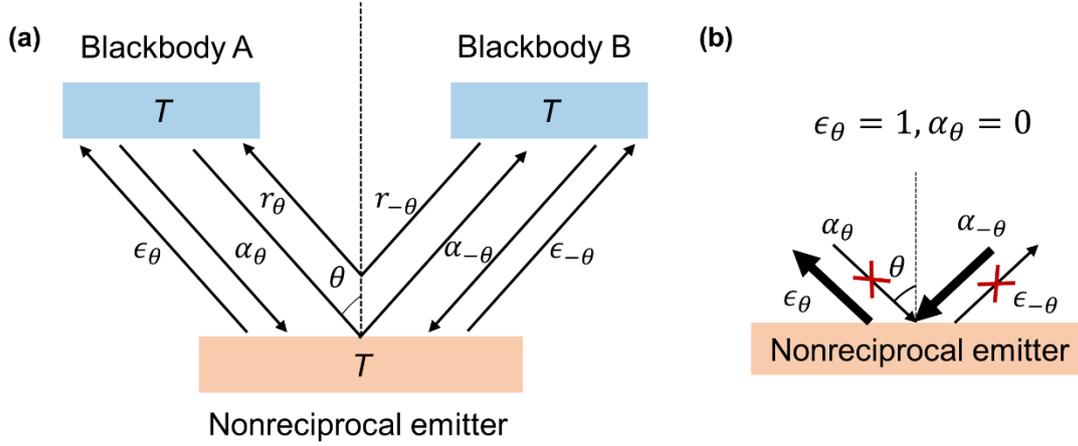

Fig. S1 (a) Energy flow diagrams in the case of a nonreciprocal radiator; (b) Schematic diagram when $\epsilon_\theta = 1$ and $\alpha_\theta = 0$.

As shown in **Fig. S1(a)**, the model consists of symmetric blackbodies A and B with a nonreciprocal emitter in between. The emitter only exchanges radiation with two independent blackbodies through two radiation channels $\theta$ and $-\theta$. Part of the radiation emitted from black body A or black body B to the emitter is absorbed, expressed by the absorptivity $\alpha_\theta$ and $\alpha_{-\theta}$, respectively. Part of radiation that is not absorbed is reflected through radiation channels, denoted by reflectance $r_\theta$ and $r_{-\theta}$, respectively. At the same time, the emitter also emits radiation to the blackbody A and B, denoted by the emissivity $\epsilon_\theta$ and $\epsilon_{-\theta}$, respectively. At this point, consider the emitter and the blackbodies in equilibrium at the same temperature T. According to the second law of thermodynamics, there is no net energy flowing into or out of the emitter, whether or not it is reciprocal. For blackbody A,

$$\epsilon_\theta + r_{-\theta} = \alpha_\theta + r_\theta = 1, \tag{1}$$

$$\epsilon_\theta - \alpha_\theta = r_\theta - r_{-\theta}. \tag{2}$$

Similarly, for blackbody B,

$$\epsilon_{-\theta} + r_\theta = \alpha_{-\theta} + r_{-\theta} = 1, \tag{3}$$

$$\epsilon_{-\theta} - \alpha_{-\theta} = r_{-\theta} - r_\theta. \tag{4}$$

Since the radiator is nonreciprocal, $r_\theta \neq r_{-\theta}$, that is, $\epsilon_{\pm\theta} \neq \alpha_{\pm\theta}$. In addition, we also can obtain $\epsilon_{\pm\theta} = \alpha_{\mp\theta}$.

**Fig. S1(b)** is the schematic diagram when $\epsilon_\theta = 1$ and $\alpha_\theta = 0$, which means that the left half space is only emission, and the right half space is only absorption. In this work, the absorptivity and emissivity at nonreciprocal bands are set as this limit state.

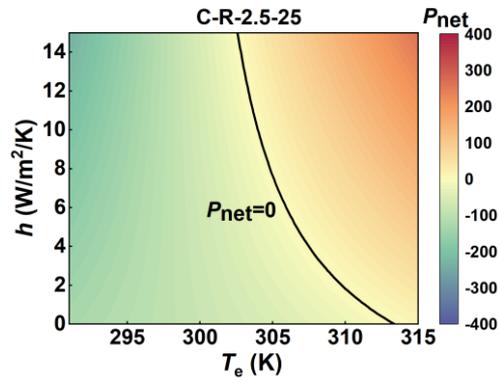

**Fig. S2** Relationship of $P_{net}$ with $T_e$ and $h$ for C-R-2.5-25 with a deep magenta color.

**Fig. S2** shows the relationship of $P_{net}$ of C-R-2.5-25 (deep magenta) with $T_e$ and $h$. It can be seen that when $P_{net} > 0$, $T_e$ is higher than 300 K. However, as can be seen from Fig. 3(h) and Table 2, since $T_e$ of C-R-2.5-25 is higher than 289.53 K, both N-4-8 and N-13-25 cannot improve the RC effect.

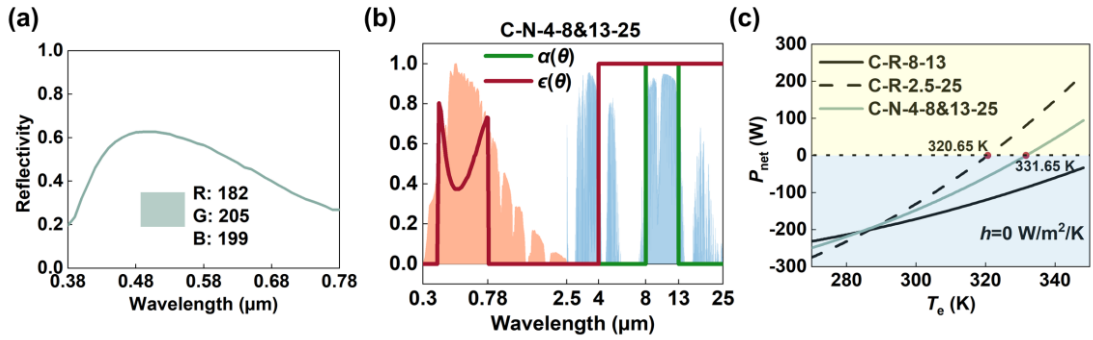

**Fig. S3** (a) Reflectance spectrum for another dark case (gray); (b) The spectrum of the nonreciprocal color radiator (C-N-4-8&13-25); (c) Net cooling power of C-R-8-13, C-R-2.5-25 and C-N-4-8&13-25 with $h$=0 W/m²/K for the gray case.

**Fig. S3** shows another dark case (gray). Fig. S3(a) is the reflection spectrum of the corresponding color, Fig. S3(b) is the corresponding nonreciprocal spectrum, and Fig. S3(c) shows the net cooling power of the nonreciprocal color radiator (C-N-4-8&13-25) and the other two types of reciprocal radiator (C-R-8-13 and C-R-2.5-25) as a function of the radiator temperature. As shown in Fig. S3(c), the conclusions are consistent with those of the dark magenta case.

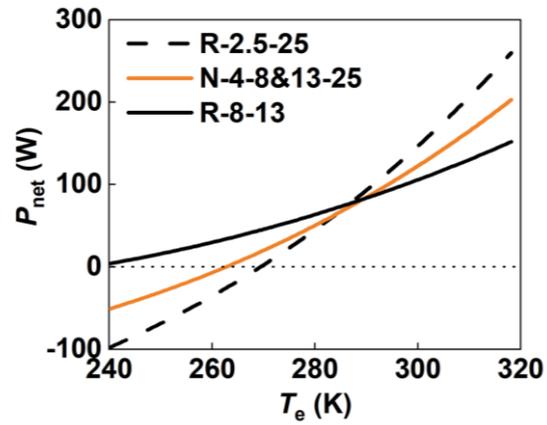

**Fig. S4** Net cooling power of R-2.5-25, R-8-13 and N-4-8&13-25 with $h$=0 W/m$^2$/K for the white case.

**Fig. S4** shows the net cooling power of R-2.5-25, R-8-13 and N-4-8&13-25 with $h$=0 W/m$^2$/K for the white case. The performance of nonreciprocal thermal radiator (N-4-8&13-25) is better than that of non-selective thermal radiator (R-2.5-25) and weaker than that of selective radiator (R-8-13), which is consistent with sections 3.1 and 3.2.